\documentclass{aa}  
\usepackage{graphicx}
\usepackage{txfonts}

\usepackage{color, colortbl}
\usepackage[hidelinks]{hyperref}
\usepackage{xcolor}
\hypersetup{
    colorlinks = true, 
    linkcolor={red!50!black},
    citecolor={blue!50!black},
    urlcolor={blue!80!black}
}
\usepackage{amstext}

\begin{document} 

\definecolor{bubbles}{rgb}{0.91, 1.0, 1.0}
\definecolor{columbiablue}{rgb}{0.61, 0.87, 1.0}
\definecolor{cream}{rgb}{1.0, 0.99, 0.82}
\definecolor{lightblue}{rgb}{0.68, 0.85, 0.9}
\definecolor{lightcyan}{rgb}{0.88, 1.0, 1.0}

   \title{Toward more accurate RR Lyrae metallicities}
   \author{Geza Kovacs
          \inst{}
          and 
          Johanna Jurcsik
          \inst{}
          }

   \institute{Konkoly Observatory, Research Center for Astronomy and Earth Sciences, 
              E\"otv\"os Lor\'and Research Network \\ 
	      Budapest, 1121 Konkoly Thege ut. 15-17, Hungary\\
              \email{kovacs@konkoly.hu}
             }

   \date{Received 09-05-2023 / Accepted DD-MM-2023}


%
%
  \abstract
{By using a large sample of published spectroscopic iron abundances, 
we point out the importance of gravity correction in deriving more 
accurate metal abundances for RR~Lyrae stars. For the $197$ stars 
with multiple spectra we find overall [Fe/H] standard deviations of 
$0.167$ (as published), $0.145$ (shifted by data source zero points) 
and $0.121$ (both zero point shifted and gravity-corrected). These 
improvements are significant at the $\sim 2\sigma$ level at each 
correction step, leading to a clearly significant improvement after 
both corrections applied. The higher quality of the gravity-corrected 
metallicities is strongly supported also by the tighter correlation 
with the metallicities predicted from the period and Fourier phase 
$\varphi_{31}$. This work underlines the need for using some external 
estimates of the temporal gravity in the chemical abundance analysis 
rather than relying on a full-fetched spectrum fit that leads to large 
correlated errors in the estimated parameters.}


\titlerunning{More accurate RR Lyrae metallicities}
\authorrunning{Kovacs \& Jurcsik}

   \maketitle
%
%
%
\section{Introduction}
\label{sect:intro}
Heavy elements -- primarily [Fe/H] but also $\alpha$-elements -- 
play a leading role in the evolution of RR Lyrae stars and also 
in their applicability as distance indicators and Galactic structure 
tracers. Unfortunately, they also have highly variable envelope/atmosphere, 
that makes the applicability of standard spectroscopic methods for 
chemical abundance analysis rather sensitive to the pulsation phase 
at the moment of the observation. 

In general, spectroscopic abundances are determined by direct model 
(or spectral library/template) fits, by adjusting basic atmospheric 
parameters, such as effective temperature $T_{eff}$, temporal gravity 
$\log g$ and turbulent velocity $V_t$, together with the assumed element 
distribution \citep[see, e.g.][for an overview of the methodologies 
used to produce today's ``industrial'' stellar abundances]{jofre2019}. 
Unfortunately, this process is rather ill-conditioned, 
and the parameters fitted are both correlated and erroneous. To improve 
the conditioning, when possible, some of the parameters (usually $T_{eff}$ 
and $\log g$) are determined externally, by using some reliable 
parameter source (e.g., spectral energy density fit from multi-waveband 
observations to derive $T_{eff}$). The gravity parameter is usually 
the most difficult parameter to estimate. 
However, extrasolar planet host stars (or the primary components  
of binaries with low-mass secondary components) and pulsating 
variables are excellent candidates for giving good estimates on 
the temporal value of $\log g$. For extrasolar planet host stars the 
gravity is well estimated from the combination of the orbital 
and stellar evolution model analyses \citep[][]{noyes2008}. For 
radially pulsating stars the temporal gravity is very closely 
estimated from the period \citep[e.g., ][]{gough1965, dekany2008} 
and from the radial velocity curve \citep[e.g., ][]{clementini1995}.  

Considering only RRab (fundamental mode) variables, here we examine the 
effect of employing a simple post-correction method on the already 
determined spectroscopic metallicities to improve the accuracy of the 
final abundances. In addition to the improvement of the internal accuracy, 
external relations are also used to test the quality of the derived 
abundances.
 
Reference to the Appendices for some additional details of the analysis 
is implicit throughout the paper. Extended materials related to this 
paper are deposited at the CDS 
site.\footnote{Metallicities and light curve parameters used in this 
paper are available at the CDS via 
http://cdsarc.u-strasbg.fr/viz-bin/qcat?J/A+A/3digitVol/Apagenumber}

%
%
\section{Method}
\label{sect:method}
The main issue in the derivation of stellar parameters on purely 
spectroscopic basis is that, in general, these parameters are 
difficult to disentangle. The success of the process depends on 
many factors and the verification of the reliability of the 
parameters derived are often presumptional (e.g., cluster's 
chemical homogeneity). For dynamical atmospheres the situation is 
coupled with changing stellar atmospheric conditions and physical 
assumptions such as the validity of local thermodinamical equilibrium. 

Here we focus on the single problem of temporal gravity. The generally 
employed spectroscopic element determination methods make this parameter 
free-floating 
\citep[together with the iron abundance and $T_{eff}$ -- see, e.g.][]{crestani2021}. 
However, for classical radial pulsators we have reliable 
estimates both for the dynamical part of the gravity (from the radial 
velocity) and also for the static part (from the pulsation equation). 
Therefore, in principle, one could use a fairly accurate estimate for 
the temporal value of the gravity, and determine only $T_{eff}$ and the 
chemical abundances. Except for \cite{clementini1995} and \cite{lambert1996}, 
we could not find any other work following this methodology. Consequently, 
we have to resort to some method that uses the already published 
atmospheric parameters and transforms the temporarily given abundances 
to the value corresponding to the static gravity. 

Our approach is purely empirical. After experiencing with various data 
sets, we found that a linear transformation applied on the published 
(spectroscopic) [Fe/H] values yields more stable [Fe/H] values and 
therefore, supposedly a better estimate of [Fe/H]\footnote{An early 
reference to the existence of such a relation can be found in 
\cite{lambert1996}. See Sect.~\ref{sect:Cg} for further discussion 
of their result in the context of this paper.} 
%
%
\begin{eqnarray}
\label{feh_logg}
{\rm [Fe/H]}_0 = {\rm [Fe/H]}_{sp} + C_g(\log g - \log g_{sp}) \hspace{2mm} ,
\end{eqnarray}
where [Fe/H]$_0$ and [Fe/H]$_{sp}$, respectively, stand for the 
transformed and the direct spectroscopic abundances (with the 
associated static and spectroscopic gravities $\log g$ and $\log g_{sp}$). 
The gravity factor $C_g$ is kept constant at the value to be determined 
by two different criteria (both seem to prefer a value near 
$0.3$ -- see Sect.~\ref{sect:Cg}).  

The static $\log g$ can be easily obtained from linear stellar pulsation 
models. Although the single parameter (i.e., period) dependence of the 
gravity is very strong, the final formula also depends on the parameter 
coverage of the models. The following formula was derived from the 
models of \cite{kovacs2021b} covering the RR~Lyrae parameter space and 
combining solar-scaled models with overall heavy element contents of 
$Z=0.0008$ and $0.001$ 
%
%
\begin{eqnarray}
\label{logp_logg}
\log g = 2.48 - 1.27\log P_0 \hspace{2mm} .
\end{eqnarray}
This formula fits the model logg values with $\sigma=0.028$, appropriate 
for the purpose of correcting the gravity effect in [Fe/H]. 
  
In addition to the gravity effect, different metallicity surveys do not 
use the very same methodology, including codes, spectral line list, 
turbulent velocity, temperature scale, solar metallicity, etc. As a 
result, there are author/source-dependent overall zero point (ZP) 
differences among the different studies. Furthermore, Eq.~\ref{feh_logg} 
is intended only to minimize the gravity dependence, but this does not 
guarantee that the resulting [Fe/H] is also close enough to the static 
value. To consider both of these effects, we employ source-by-source 
differential corrections in the following sense  

%
%
\begin{eqnarray}
\label{feh_zp}
[Fe/H](i) = [Fe/H]_0(i) + DZP(i) \hspace{2mm} ,
\end{eqnarray}
where [Fe/H]$_0(i)$ is the gravity-corrected abundance for 
some star in the $i$-th source. Because the methodology followed 
by \cite{clementini1995} and \cite{lambert1996} seems to be the 
closest to the type of method we advocate, the DZP-s are determined 
in a way which considers the averages of all available metallicities, 
but these are additionally shifted (uniformly) by an amount of 
$-0.08$~dex that results the smallest ZP corrections for the 
metallicities from the above two 
publications.\footnote{Even though these two publications 
are based on very close principles, the resulting abundances 
for the three common stars differ by $0.1$--$0.2$~dex. This paper 
uses the average of the Fe I and Fe II values in the `photometric' 
section of Table~3 from \cite{lambert1996} and Table~12 of 
\cite{clementini1995}.}

To sum it up, for any {\em fixed} $C_g$, the algorithm of transforming 
the published spectroscopic abundances to a uniform scale that observes 
the start-by-star $\log g$ and the survey-by-survey ZP dependences, 
constitutes the following steps: 

\begin{itemize}
\item[1.]
{\em Gravity correction:} 
use Eqs.~\ref{feh_logg},\ref{logp_logg} for all stars in all sources. 
\item[2.]
{\em Estimate the `ridge'}:
use each star with multiple sources and compute simple averages 
from the logg-corrected [Fe/H]$_0$ values. 
\item[3.]
{\em ZP corrections:}
Compute DZP for each source as given by Eq.~\ref{feh_zp} by considering 
the average difference between the ridge and the source values. 
Add $-0.08$~dex as a global correction to each DZP. 
\item[4.]
{\em Final [Fe/H]:}
Loop back to step 1., and derive the final metallicities FEH by employing 
the gravity correction on the ZP-corrected input [Fe/H]$_{sp}$ values and 
perform simple averaging for objects with multiple measurements. 
\end{itemize}

The above core scheme is run for the scanned $C_g$ values in the 
process to be aimed at the optimization of $C_g$ by using various 
minimization criteria for the RMS of the resulting metallicities (see 
Sect.~\ref{sect:Cg}).

%
%
\section{Data sets}
\label{sect:data}
In gathering high-quality [Fe/H] data, first we made a search for 
publications with simultaneous $T_{eff}$, $\log g$ data based on high 
dispersion spectroscopic (HDS) observations. Then, these data were 
further extended by big survey data (LAMOST and GALAH). 
Although LAMOST does not use a HDS equipment, the method 
employed is very similar to those of the HDS small-scale studies. 
Therefore, for simplicity, we label these data also as HDS. The 
final list of HDS data is based on an additional cross-check of 
the above big survey databases with the list of \cite{muraveva2018}. 
We recall that only RRab variables are included in all these and 
subsequent data sets.  

%
%
\begin{table}[h]
\centering
\begin{minipage}{200mm}
\caption{HDS inventory}
\label{hds_inv}
\scalebox{1.0}{
\begin{tabular}{lrrrcc}
\hline
 Source & $N_{obj}$ & $N_{sp}$  &  DZP  &  RMS & $N_{clip}$\\
\hline\hline
ta22 &  21 &  31 & $-0.1053$ & 0.0596 &  0\\
cr21 & 121 & 184 & $-0.0511$ & 0.0808 &  2\\
la96 &  15 &  15 & $ 0.0572$ & 0.0466 &  0\\
li13 &  22 &  33 & $-0.1654$ & 0.0497 &  0\\
ch17 &  28 &  70 & $-0.1383$ & 0.0998 &  0\\
cl95 &  10 &  10 & $-0.0471$ & 0.0966 &  0\\
fo11 &  11 & 165 & $-0.0601$ & 0.0193 &  0\\
ne13 &  24 &  24 & $-0.1376$ & 0.1175 &  0\\
lam5 & 119 & 216 & $-0.0984$ & 0.0798 &  0\\
lal5 &  81 & 133 & $-0.0395$ & 0.0761 &  1\\
gala &  22 &  22 & $-0.1629$ & 0.0893 &  0\\
an18 &  26 &  51 & $-0.0852$ & 0.0882 &  0\\
pa15 &  18 &  54 & $-0.1378$ & 0.1077 &  0\\
an21 &   7 &   7 & $-0.1769$ & 0.0838 &  0\\
fe96 &   9 &   9 & $ 0.0942$ & 0.1085 &  0\\
fe97 &  38 &  38 & $-0.0808$ & 0.1495 &  1\\
so97 &  23 &  47 & $-0.1349$ & 0.1235 &  2\\
\hline
Total: &  269 & 1109 & --- & --- &    6\\
\hline
\end{tabular}}
\end{minipage}
\begin{flushleft}
\vspace{-5pt}
{\bf Notes:}
$N_{obj}=$~number of objects; $N_{sp}=$~number of spectra;
DZP$=$~zero point shift for [Fe/H] (including differential 
and global shifts -- see Sect.~\ref{sect:method}); 
RMS$=$~standard deviation of the final metallicities for 
the residuals between the source items and the averages of 
the values obtained from all sources with multiple spectra; 
$N_{clip}=$~Number of $3\sigma$-clipped objects.\\
{\bf References:} 
ta22=\cite{takeda2022}; 
cr21=\cite{crestani2021};  
la96=\cite{lambert1996};  
li13=\cite{liu2013}; 
ch17=\cite{chadid2017}; 
cl95=\cite{clementini1995}; 
fo11=\cite{for2011}; 
ne13=\cite{nemec2013}; 
lam5=\cite{xiang2019}; 
lal5=\cite{luo2019a,luo2019b}; 
gala=\cite{buder2021}; 
an18=\cite{andrievsky2018}; 
pa15=\cite{pancino2015}; 
an21=\cite{andrievsky2021}; 
fe96=\cite{fernley1996}; 
fe97=\cite{fernley1997}; 
so97=\cite{solano1997} 
\end{flushleft}
\end{table}
Some details of the HDS inventory used are shown in Table~\ref{hds_inv}. 
We also checked two more sources: \cite{sprague2022}, based on the 
APOGEE survey and \cite{gilligan2021}, from the observations made 
by SALT/HRS. Unfortunately, both of these sets proved to be too 
noisy with respect of the sources listed in Table~\ref{hds_inv}. 

\begin{figure}[h]
\centering
\includegraphics[width=0.48\textwidth]{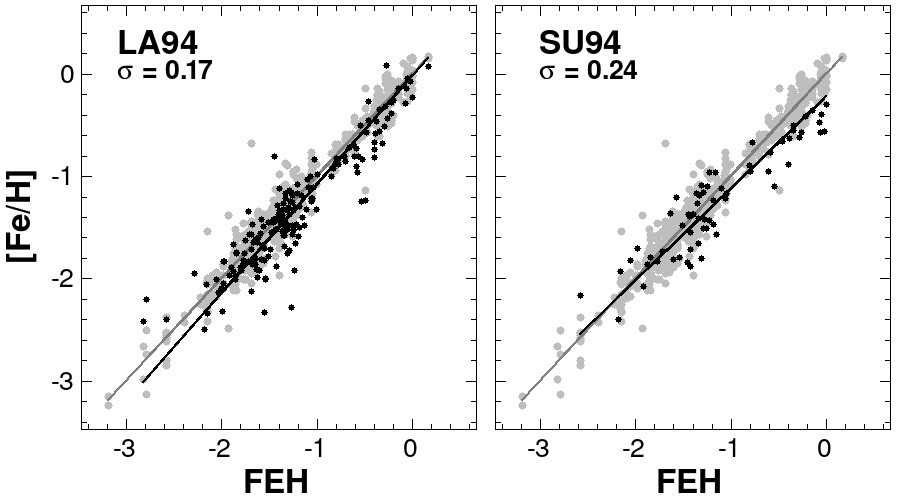}
\caption{HDS ridge metallicities (FEH) vs individual HDS (gray dots) 
         and LDS (black dots) metallicities. The corresponding 
	 regression lines are shown by gray and black lines. The 
	 standard deviations of the LDS regressions are shown in 
	 the upper left corners of the corresponding LDS sources:  
	 LA94 for \cite{layden1994} and SU94 for \cite{suntzeff1994}.} 
\label{lds_vs_hds}
\end{figure}

As a consistency check, we also examined the earlier low dispersion 
spectroscopic (LDS) data by \cite{layden1994} and \cite{suntzeff1994} 
(hereafter LA94 and SU94, respectively). We recall that the LDS 
metallicities are based on the calibration of spectral indices (or 
equivalent widths) that are conveniently measured also on low 
dispersion spectra. These data are then calibrated on proper sets 
of [Fe/H] derived from HDS spectra (with the HDS methodology). Once 
that LDS parameters have been calibrated, they can be quickly 
applied to determine the metallicities of new objects, often not 
accessible by HDS instruments.    
Somewhat surprisingly, we 
found that these early sets follow quite well the HDS ridge 
(star-by-star average [Fe/H], see Fig.~\ref{lds_vs_hds}). Both sets 
show systematic differences with respect to the HDS values. We 
approximate these trends by straight lines. The regression coefficients 
are given in Table~\ref{lds_reg}. The improved quality of the merged 
(HDS \& LDS) data will be demonstrated in Sect.~\ref{sect:jk96}.  

Concerning the calibration of the gravity- and source ZP-corrected 
[Fe/H] relative to the Sun, we note that an overwhelming majority 
of the sources use $\log \epsilon(Fe) = 7.50$. Therefore, [Fe/H] 
presented in this work is considered to be tied to this solar value. 

%
%
\begin{table}[h]
\centering
\begin{minipage}{200mm}
\caption{LDS to HDS transformations}
\label{lds_reg}
\scalebox{1.0}{
\begin{tabular}{lcccrr}
\hline
 Source & $c_1$ & $c_2$  &  RMS & $N$ & $N_{clip}$\\
\hline\hline
LA94 & 0.0216 & 0.9433 & 0.1740 & 184 &  8\\
SU94 & 0.2514 & 1.1147 & 0.2359 &  64 &  0\\
\hline
\end{tabular}}
\end{minipage}
\begin{flushleft}
\vspace{-5pt}
{\bf Notes:} [Fe/H]$_{HDS} \sim c_1 + c_2$[Fe/H]$_{LDS}$; 
the fit was made to FEH (gravity- and source ZP-corrected 
HDS [Fe/H]); $N_{clip}=$~number of $3\sigma$-clipped objects. 
See accompanying Fig.~\ref{lds_vs_hds}.    
\end{flushleft}
\end{table}
%

%
%
%
%

%
%
\section{The gravity factor}
\label{sect:Cg}
\begin{figure}[h]
\centering
\includegraphics[width=0.40\textwidth]{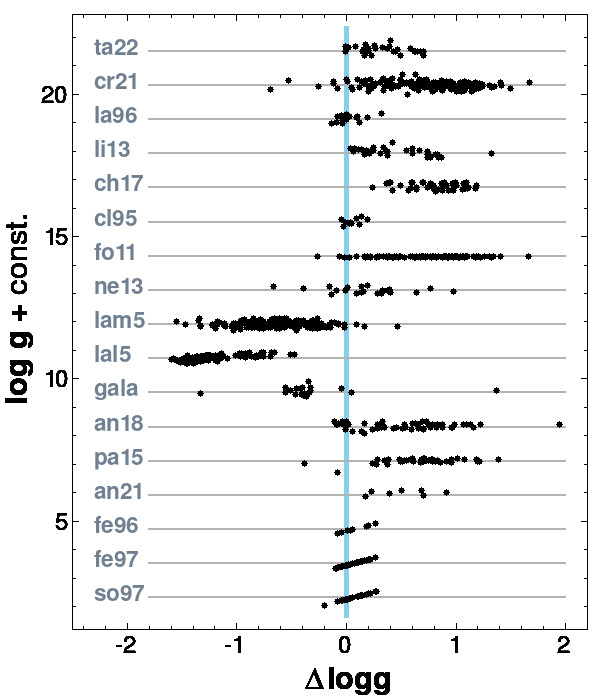}
\caption{Gravity differences (static value, $\log g$, from 
         Eq.~\ref{logp_logg} minus published spectroscopic value) 
	 for the individual objects of the different sources. 
	 For each source, a uniform vertical shift was applied 
	 to separate it from the data of other sources. See 
	 Table~\ref{hds_inv} for the source acronyms.} 
\label{check_logg}
\end{figure}

Before dwelling more deeply into the gravity correction, it is 
worth examining the distribution of the published gravities of the 
various surveys and targeted studies. Figure~\ref{check_logg} shows 
the distribution of the gravity differences  
$\Delta \log g = \log g - \log g_{sp}$ for the various sources. 
Large deviations can be seen in both directions, suggesting 
improper pulsation phasing of the observations, or, more likely, 
large errors in the spectroscopic gravities. The surprisingly 
flat pattern of $\log g$ for the set of {\em fo11} \citep[][]{for2011} 
is due to their specific sample of stars, with periods concentrated 
in a narrow range. The striking linear dependence for {\em fe96}, 
{\em fe97} and {\em so96} 
\citep[respectively,][]{fernley1996,fernley1997,solano1997} is due 
to the fixed $\log g_{sp}$ values in their analysis. The 
$\Delta \log g$ ranges are particular small and quite close to 
the values predicted by Eq.~\ref{logp_logg} for {\em la96} and 
{\em cl95} \citep[respectively,][]{lambert1996,clementini1995}, 
indicating proper phasing of the observations in these studies. 

%
%
\begin{figure}[h]
\centering
\includegraphics[width=0.40\textwidth]{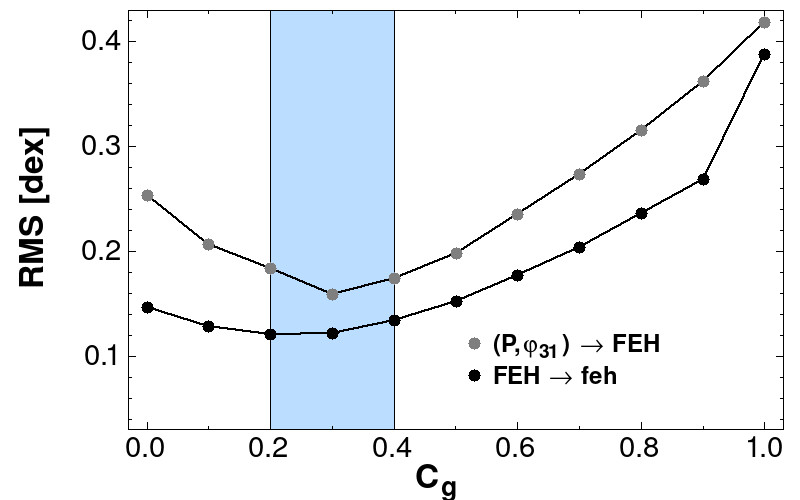}
\caption{Dependence of the fit RMS values on the gravity correction 
         factor $C_g$ (see Eq.~\ref{feh_logg}). The plot shows the 
	 scans resulting from the direct estimates of the scatter 
	 of the individual transformed abundances \{feh\} around the 
	 average values \{FEH\} and the variation of the fit 
	 RMS of the Fourier-based estimates. The shaded box indicates 
	 the optimum regime for $C_g$.} 
\label{cgg_scan}
\end{figure}

After a detailed examination of the star-by-star dependence of 
the published [Fe/H] values on gravity difference 
$\log g - \log g_{sp}$, we found that the gravity factor $C_g$ 
(Eq.~\ref{feh_logg}) has a visible target dependence. For instance, 
Z~Mic has remarkably constant [Fe/H] implying $C_g\sim 0.0$, 
whereas V1645~Sgr is best fitted by a steep $C_g$ of $\sim 0.6$. 
However, using individual $C_g$ would lead to a high level of 
overfitting, due to the modest number of spectra for most of the 
stars.\footnote{Nevertheless, for further support of our adopted 
value of $C_g$, it is useful to examine the distribution of the 
star-by-star fitted $C_g$ values and check the degree of spread 
of these values. See Appendix~\ref{app_C} for this test.} 
Therefore, we decided to search for the best global value 
of $C_g$, by minimizing the scatter of the $\log g$-transformed 
and source ZP-adjusted abundances around the average values for 
all stars with multiple spectra ($197$ stars and $1037$ spectra 
altogether). Black dots in Fig.~\ref{cgg_scan} show the run of 
RMS (individual transformed metallicities \{feh\} minus their 
average FEH) for the full scan range of $C_g$. A preference for 
$C_g\sim 0.2$ is clearly visible. 

Yet another way to optimize $C_g$, is to scan the dependence of the 
RMS of the $(P,\varphi_{31})\rightarrow$~FEH fit \citep{jurcsik1996}, 
this time using the full HDS set, including single spectra data 
(263 objects with available $V$ light curves -- see Sect.~\ref{sect:jk96}). 
Gray dots in Fig.~\ref{cgg_scan} indicate a preference of high 
significance for a global value of $C_g\sim 0.3$. 

%
%
To further constrain $C_g$, we also attempted to test the tightness 
of the [Fe/H]$-M_V$ \citep[e.g.,][]{castellani1991} and the near 
infrared period-luminosity-metallicity \citep[PLZ, i.e.,][]{bono2003} 
relations. Unfortunately, both of these relations have relatively weak   
metal dependences. In addition, this dependence is further masked by 
noise, internal physical scatter (for the [Fe/H]$-M_V$ relation) and 
the correlation between the period and metallicity (for the PLZ test). 
Therefore, we rely only on the direct tests shown in Fig.~\ref{cgg_scan}. 
These tests suggest values of $C_g\sim0.25$, which we will use 
throughout the paper. 

\begin{figure}[h]
\centering
\includegraphics[width=0.50\textwidth]{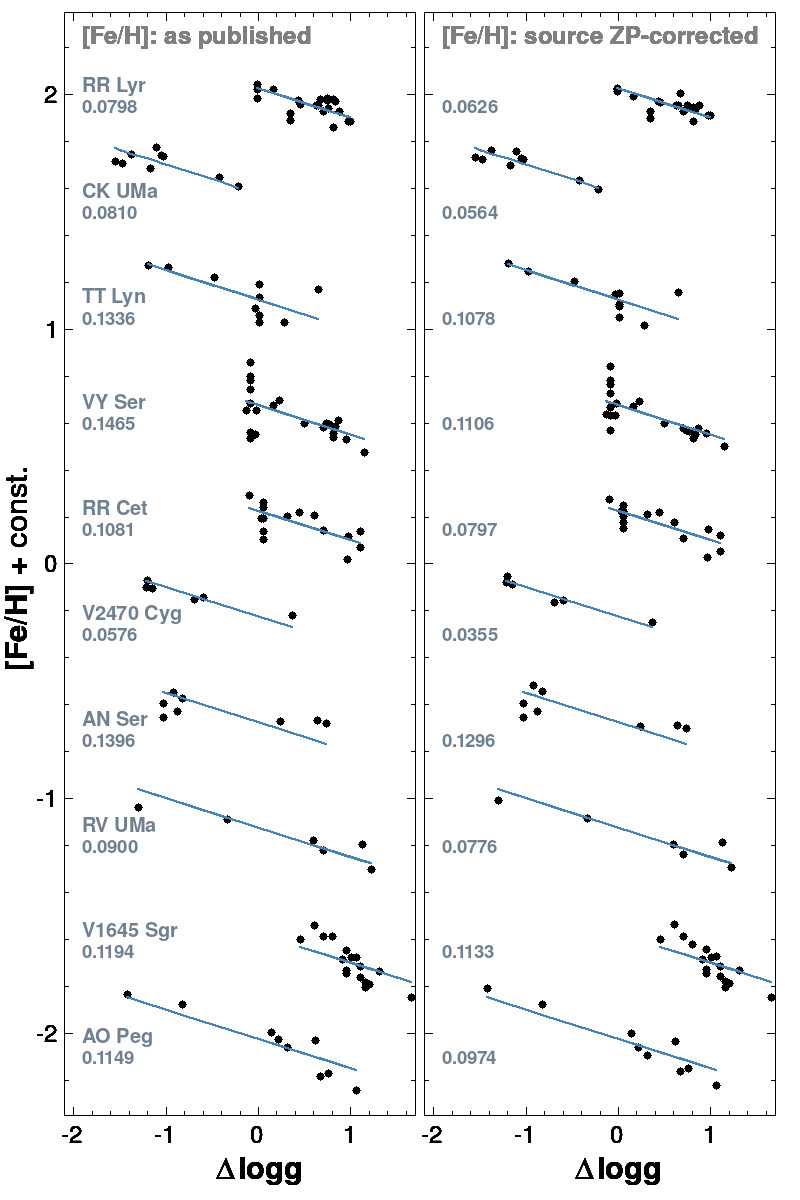}
\caption{Examples of the star-by-star correlation between the 
         gravity difference and the published abundances (left panel). 
	 The same correlation is shown in the right panel after 
	 source ZP corrections. For an easier visualization, vertical 
	 shifts were applied together with a factor of two decrease 
	 of the [Fe/H] ranges (including the slope of the fitted 
	 straight line with an original value of $-0.25$). Numbers 
	 at the individual objects show the RMS values around the 
	 straight lines). See text for further details.} 
\label{logg_feh_10}
\end{figure}

For a very simple demonstration of the differential 
$\log g$-dependence of the published spectroscopic abundances, 
in Fig.~\ref{logg_feh_10} we plot these [Fe/H] values as a function 
of $\Delta \log g = \log g - \log g_{sp}$ (with $\log g$ as given 
by Eq.~\ref{logp_logg}). In the left panel, all [Fe/H] were 
taken straight from the respective publications and plotted 
against $\Delta \log g$. In the right panel we employed 
source ZP corrections as described in Sect.~\ref{sect:method} 
and presented in Table~\ref{hds_inv}. For each star, the [Fe/H]$_{sp}$ 
values were fitted by a straight line of Eq.~\ref{feh_logg} with 
$C_g=0.25$ and [Fe/H]$_0$ adjusted by a least-squares fit of equal 
weights. 

As noted earlier, although in most cases the adopted slope 
is in good agreement with the observations, there are cases 
with considerable differences (steeper, like V1645~Sgr and 
shallower, such as AN~Ser). Source ZP corrections further 
improve the fit as can be seen by the insets at each object 
showing the RMS of the residuals around the straight lines.

%
%
%

%
%
%
\begin{figure}[h]
\centering
\includegraphics[width=0.50\textwidth]{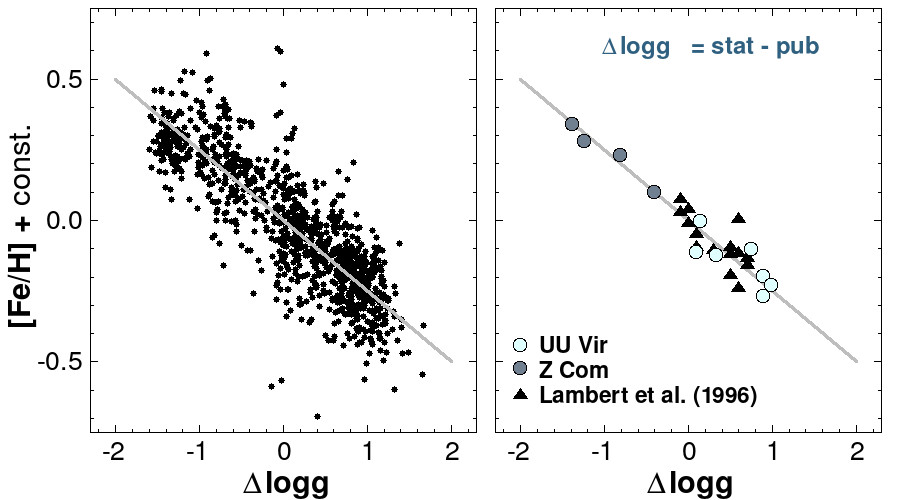}
\caption{{\em Left:} Correlation between the published HDS 
         metallicities and the difference between the static 
	 and spectroscopic gravities for the $197$ 
	 stars with multiple spectra. Vertical shifts were 
	 applied according to Eq.~\ref{feh_logg} to avoid 
	 scatter due to differences in the stellar metallicities. 
	 {\em Right:} As in the left panel, 
	 but only for two well-fit examples. Triangles are for 
	 the RRab sample from Table~3 of \cite{lambert1996} 
	 by using the `photometric' and `spectroscopic' results 
	 (see text for further details).} 
\label{gg_all}
\end{figure}

We can also visualize the $\Delta \log g \rightarrow$[Fe/H] 
correlation by collapsing the data in the vertical axis (i.e., 
not applying star-by-star vertical shifts). The left panel of 
Figure~\ref{gg_all} shows the result of this type of data handling 
by using the full sample of $197$ stars with multiple measurements. 
In the right panel, in addition to showing two well-fit cases we 
also exhibit the specific data from \cite{lambert1996}. To the best 
of our knowledge their work is the only one (in the context 
of RR~Lyrae stars) comparing metallicities from the traditional 
spectroscopic method and those obtained by using known values of 
the temporal gravity and theoretical spectral models. 

Using Table~3 of \cite{lambert1996}, first we calculated the average 
of the `photometric' metallicities derived from the FeI and FeII 
lines.\footnote{Although the correlation with $\Delta \log g$ is 
tighter for FeII \citep[][]{lambert1996}, we used the average for 
consistency with the final data set of this paper.}   
These metallicities were obtained by employing the $\log g$ values 
mostly from various Baade-Wesselink analyses, and, in fewer cases, 
from narrow-band photometry. Then, we used the `spectroscopic' 
metallicities\footnote{These are derived from direct theoretical 
spectrum fits, based on equating the abundances obtained from spectral 
lines of various ionization levels. The gravity is computed as part of 
the fitting process.} to calculate the difference between the these 
and the `photometric' metallicities. The corresponding gravity differences 
(`spectroscopic' minus `photometric') are also listed in the same 
table of \cite{lambert1996}. Except for a vertical adjustment of 
$-0.09$~dex, we plotted straight these published values in Fig.~\ref{gg_all}. 
The $15$ RRab stars follow remarkably well the trend observed by 
the other two individual objects with multiple measurements.    

Although it is not easy to find a single source for the 
correlation between $\Delta \log g$ and [Fe/H], non-LTE effects 
are likely play an important role \citep{lambert1996}. The work 
by \cite{luck1985} on intermediate-mass supergiants (including 
Cepheids) was the first to point out the systematic differences 
between the gravities obtained from the spectroscopic analyses 
and pulsation models. Being less sensitive to non-LTE effects, 
\cite{kovtyukh1999} suggest using Fe~II lines for abundance and 
gravity estimates in the case of Cepheids. \cite{lambert1996} 
lend on a similar conclusion for RR~Lyrae stars.     

\begin{figure}[h]
\centering
\includegraphics[width=0.40\textwidth]{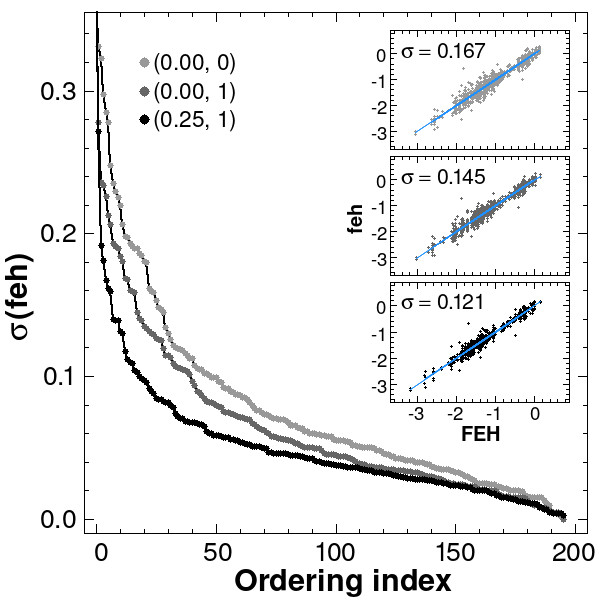}
\caption{Ordered standard deviations of the multiple [Fe/H] 
         values under various data handling as given by  
	 the code list in the upper left part of the figure. 
	 First number in the code symbol corresponds to the 
	 logg correction term ($C_g$ in Eq.~\ref{feh_logg}), 
	 whereas the second number indicates if source-by-source 
	 zero point correction was (1) or was not (0) employed. 
	 The inserted panels show the scatter of the star-by-star 
	 individual metallicity values \{feh\} around the 
	 corresponding average metallicities FEH. The equality 
	 lines are indicated by light blue. The standard deviations  
	 are shown by the internal labels.} 
\label{feh_sig_3}
\end{figure}

To visualize the improvement of the quality of the derived 
metallicities at various stages of the process, in Fig.~\ref{feh_sig_3} 
we show the star-by-star standard deviations for the above set of 197 stars. 
The ZP correction yield an RMS improvement which is significant 
at the $2\sigma$ level. This is quite close to the additional 
$2.3$ sigma significance of the gravity correction on the 
ZP-corrected values. All these indicate essentially the same 
level of importance of both types of correction.  

%
%
%

%
%
\section{The $(P,\varphi_{31})\rightarrow$[Fe/H] fit}
\label{sect:jk96}
Here we test further the gravity- and source zero point-corrected 
metallicities FEH (Sects.~\ref{sect:Cg} and \ref{sect:data}). 
This test also casts more light on the potentials and limitations 
of the metallicity determination based on light curve analyses 
\citep[][]{kovacs1995,jurcsik1996}. Unfortunately, this relation 
still remains to be purely empirical, without any theoretical foundation 
\citep[except perhaps for the work of][]{feuchtinger1999}, in spite 
of the practical applicability and current recalibrations on 
various data sets \citep[e.g.,][]{dekany2022,li2023}.\footnote{At the 
same time, it is worthwhile to note that the large datasets used 
in this paper further strengthen the strong preference for the 
$(P,\varphi_{31})$ dependence against any other two-parameter 
combinations, including $\varphi_{41}$ and $A_{tot}$ (depending on 
the data sets, at the $1-4$ and $8-12$ $\sigma$ levels, respectively).}  

We use the light curves from the ASAS and ASAS-SN surveys 
\citep[][]{pojmanski1997, shappee2014, christy2023} and the Fourier 
decompositions presented by \cite{jurcsik1996}. Because our approach 
here is to utilize as many spectroscopic [Fe/H] as possible, we include 
{\em all RRab stars}, irrespectively if they are monomode or Blazhko. 
If the target is of this latter type, it is frequency analyzed, and the 
mid curve (i.e., that part of the light curve that comes from the monomode 
pulsation) is used to estimate the Fourier phase. Additional details 
of the light curve analysis can be found in Appendix~\ref{app_B}.

The metallicity data sets to be dealt with are listed in 
Table~\ref{data_31}. These sets were selected to investigate: 
(i) effect of merging LDS and HDS data to aim for more accurate 
test base (set A, containing stars with average metallicities 
derived from HDS and LDS sources); 
(ii) more realistic situations, when the data quality strongly changes 
within the sample (set B, with mixed items containing only either 
HDS or LDS, or type A metallicities). 
The RMS values indeed show that the merged data yield better result. 
Assuming Gaussian distributions for the residuals (an assumption that 
has only partial validity -- see later), we get that the RMS differences 
with respect to set A for A$_{HDS}$, A$_{LDS}$ and B are significant 
at the $1.7$, $2.5$ and $1.4$ sigma levels, respectively. For set 
A$_{HDS}$, employing only HDS source ZP corrections (i.e., no gravity 
correction), we get a fit with an RMS of $0.256$. With respect to set 
A, this increase is significant at the level of $5.7$ sigma.

%
%
\begin{table}[h]
\centering
\begin{minipage}{200mm}
\caption{Datasets for the $(P,\varphi_{31})\rightarrow$[Fe/H] test}
\label{data_31}
\setlength{\tabcolsep}{5pt}
\scalebox{1.0}{
\begin{tabular}{lclcc}
\hline
 Name & N$_{FEH}$ & Content  &  RMS$_{fit}$ & N$_{clip}$\\
\hline\hline
A           &  187  & HDS with LDS  & 0.163 &  \phantom{0}3\\
A$_{HDS}$   &  187  & HDS only in A & 0.186 &  \phantom{0}2\\
A$_{LDS}$   &  187  & LDS only in A & 0.199 &  \phantom{0}1\\
B           &  390  & A$+$HDS only$+$LDS only & 0.179 & 11\\
\hline
\end{tabular}}
\end{minipage}
\begin{flushleft}
\vspace{-5pt}  
\end{flushleft}
\end{table}
%

%
%
\begin{figure}[h]
\centering
\includegraphics[width=0.35\textwidth]{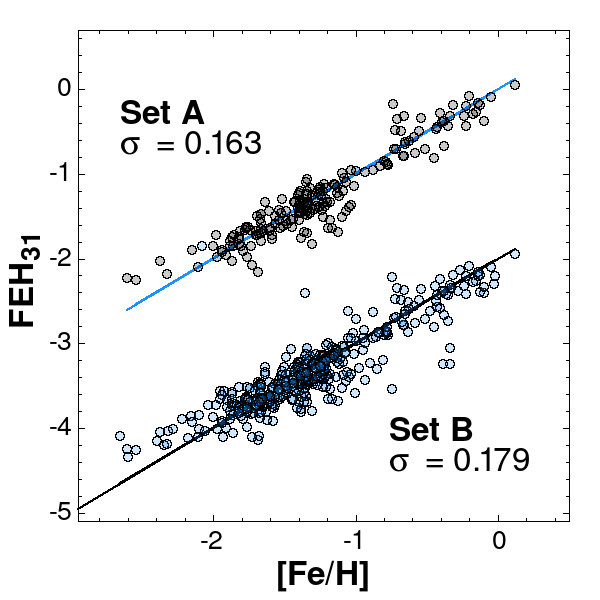}
\caption{Metallicity predictions FEH$_{31}$ from the 
         period and Fourier parameter $\varphi_{31}$. Datasets 
	 given in Table~\ref{data_31} are used including both 
	 $\log g$- and ZP-corrected HDS data and LDS data as 
	 discussed in Appendix~\ref{app_A}. These metallicities 
	 are shown on the horizontal axis. Straight lines 
	 indicate the equality values for [Fe/H]. The plot 
	 for set B has been shifted downward for better 
	 visibility.} 
\label{p_31_fit}
\end{figure}

The corresponding regressions for sets A and B are shown in 
Fig.~\ref{p_31_fit}. We recall that the samples shown include 
{\em all stars} (i.e., even those indicated as `clipped' objects in 
Table~\ref{data_31}. Systematically high-fit objects at the low 
metallicity end are clearly visible. Assuming that the Fourier 
method is sensitive to the global metallicity [M/H] rather than 
to [Fe/H], then, this effect may come from the stronger $\alpha$ 
element enhancements at lower metallicities. On the other hand, 
we see also systematically low-fit values for quite a large 
number of stars in the mid- to high-metallicity part. A brief test  
of the effect of possible low (or perhaps negative) $\alpha$ 
enhancement for these stars by using the estimates of \cite{crestani2021} 
did not suggest any statistically significant improvement. 
A more detailed examination of these stars vaguely implicated that 
the underestimation by the Fourier method may be partially attributed 
to the slight excess of Blazhko stars (however, we do not have an 
explanation why these stars may cause such an effect).     

In addition to the visual inspection of the residuals, we can 
also examine their distribution. It turns out (see Appendix~\ref{app_D}) 
that none of the data sets follow Gaussian distribution in the 
outskirt of their ordered residuals. About $20$\% of the stars 
in each sample belong to a non-Gaussian subset. Although 
several outliers can be explained by light curve or metallicity 
errors/peculiarities, both in the non-Gaussian set and also among 
the simple outliers, there are stars with excellent metallicities 
and accurate light curves. Stars such as X~Ari, AL~CMi or V0341~Aql 
remain among the puzzling details of the Fourier method for [Fe/H] 
determination. 

From purely practical point of view, a large fraction of the 
observed [Fe/H] values are fitted remarkably well with $(P,\varphi_{31})$. 
The errors are less than $0.2$~dex for over $70$\% and less than 
$0.1$~dex for about $50$\% of the stars in our samples. 

Finally, for $V$ light curves (using the monomode components for 
Blazhko stars) the updated formula derived from set A of 
Table~\ref{data_31} reads as 
follows\footnote{By using set B we get 
$FEH_{31} = -4.845 - 5.385\,P + 1.274\,\varphi_{31}$, yielding 
estimates that deviate from the values predicted by 
Eq.~\ref{eq_p_31_fit} with a standard deviation of $0.017$~dex.}
%
%
\begin{eqnarray}
\label{eq_p_31_fit}
FEH_{31} = -5.088 - 5.268\,P + 1.311\,\varphi_{31}\hspace{2mm}. 
\end{eqnarray}
%

%
%
%
\section{Conclusions}
\label{sect:conclude}
By using the best available iron abundances based on traditional 
spectral analyses of fundamental mode RR~Lyrae stars, we arrived 
to the following conclusions.
\begin{itemize}
\item[$-$]
Current spectroscopic abundances suffer from considerable scatter 
both internally (within a given survey) and externally (between 
the different surveys).
\item[$-$]
The internal scatter can be mitigated on a star-by-star basis by 
employing a linear correction including the difference   
between the static and the temporal (spectroscopic) gravities. 
\item[$-$]
The external (zero point) differences can be eliminated by 
uniform survey-by-survey shifts based on the common stars in the 
different surveys.
\item[$-$]
Both corrections yield roughly the same degree of improvement in 
the overall quality of the final abundances, resulting in a RMS 
of $0.12$~dex around the averages for stars with multiple measurements. 
This is an improvement above $4$ sigma with respect to the RMS of 
$0.17$~dex of the published (uncorrected) values.  
\end{itemize}
The relatively large differences among the spectroscopic abundances 
have led us to examine the possible utilization of the metallicity 
estimates based on spectral index methods \citep{layden1994,suntzeff1994}. 
Interestingly, we found these metallicities quite comparable with 
the overall higher accuracy direct spectroscopic metallicities. 
The merged set of $187$ stars (set A in Table~\ref{data_31}) represents 
the most accurate metallicities used in this paper. 

The higher quality [Fe/H] reflects also on the tightness of the 
relations involving [Fe/H]. Although the effect is rather mild 
on the [Fe/H]$\rightarrow M_V$ and near infrared 
$(log P, {\rm [Fe/H]})\rightarrow M_{Ks}$ relations, the improvement 
on the $(P,\varphi_{31})\rightarrow$[Fe/H] fit is far more 
significant. In this update we extended the applicability of the 
Fourier method to Blazhko stars by using the average light curves, 
as derived from the Fourier fits including also the Blazhko modulation. 
The precision of the [Fe/H] estimates are, respectively, within $0.2$ 
and $0.1$ dex for some $70$\% and $50$\% of the samples investigated.   
The Fourier-based [Fe/H] yield a little tighter correlations for the 
above luminosity relations. At the same time, we found puzzling differences 
between the Fourier estimates and the observed values in several stars. 
We cannot offer any explanation for this phenomenon at this moment, 
except perhaps for the vague idea that the Fourier estimate refers more 
likely to the overall metal abundance rather than [Fe/H] and that for 
some reasons some of the stars are $\alpha$ element deficient (rather 
than enhanced).   

Based on the improvement presented in this paper, we strongly argue 
for the reexamination of methodology of element abundance determination 
in RR~Lyrae stars, and in pulsating stars, in general. The method used 
by \cite{clementini1995} and \cite{lambert1996} are important examples  
for the path we think would be very fruitful to pursue.

%
%
\begin{acknowledgements}
%
We appreciate the discussion with Giuliana Fiorentio on the intricacy 
of $\alpha$-enhancement. 
%
It is a pleasure to thank the useful comments by Jayasinghe A. 
Tharindu on the compatibility of the ASAS and ASAS-SN light curves. 
We thank the referee for drawing our attention on some earlier 
works relevant for this paper. 
This research has made use of the VizieR catalogue access tool, CDS, 
Strasbourg, France (DOI: 10.26093/cds/vizier). 
%
This work has made use of data from the European Space Agency (ESA) 
mission {\it Gaia} (\url{https://www.cosmos.esa.int/gaia}), 
processed by the {\it Gaia} Data Processing and Analysis Consortium 
(DPAC, \url{https://www.cosmos.esa.int/web/gaia/dpac/consortium}). 
Funding for the DPAC has been provided by national institutions, 
in particular the institutions participating in the {\it Gaia} 
Multilateral Agreement.
%
This research has made use of the NASA/IPAC Infrared Science Archive, 
which is funded by the National Aeronautics and Space Administration 
and operated by the California Institute of Technology.
%
This research has made use of the International Variable Star Index 
(VSX) database, operated at AAVSO, Cambridge, Massachusetts, USA.
%
Guoshoujing Telescope (the Large Sky Area Multi-Object Fiber Spectroscopic 
Telescope LAMOST) is a National Major Scientific Project built by the 
Chinese Academy of Sciences. Funding for the project has been provided 
by the National Development and Reform Commission. LAMOST is operated 
and managed by the National Astronomical Observatories, Chinese Academy 
of Sciences.
%
This work made use of the Third Data Release of the GALAH Survey 
\citep{buder2021}. 
Supports from the National Research, Development and Innovation 
Office (grants K~129249 and NN~129075) are acknowledged. 
\end{acknowledgements}

%
%

%

%
%
%
\begin{appendix}
\section{Metallicities}
\label{app_A}
\begin{figure}[h]
\centering
\includegraphics[width=0.48\textwidth]{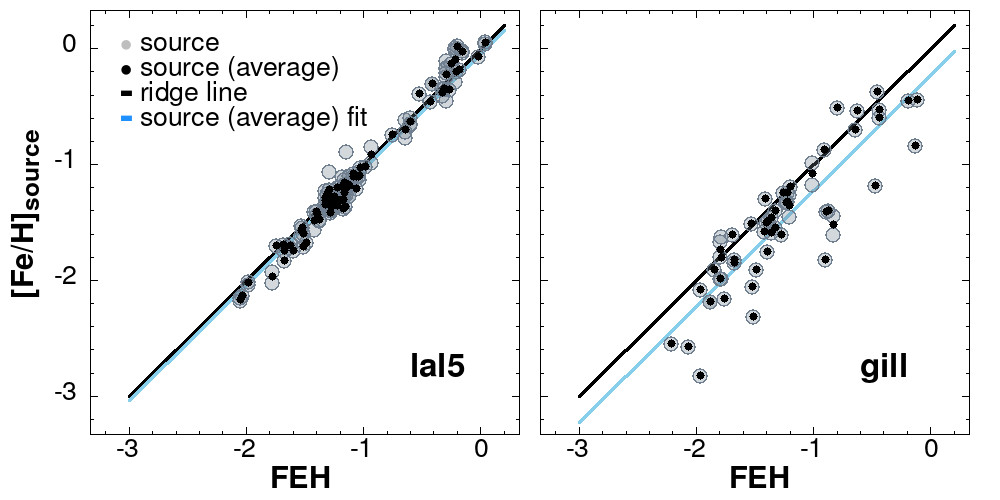}
\caption{Source-dependent individual vs ridge metallicities 
         (FEH, averages computed from all sources). Black line 
	 shows the ridge, blue line shows the fit to the particular 
	 source averages (assuming only a constant shift relative 
	 to the ridge line for a given source). Black dots are for 
	 the average metallicities, pale, larger circles are for 
	 the individual metallicities. The left and right panels 
	 show, respectively, the well-fit case of \cite{luo2019a,luo2019b} 
	 (LAMOST DR5, DR7) and the poor-fit case of \cite{gilligan2021}. 
	 The latter is not included in our final HDS sample. 
	 Table~\ref{hds_inv} is referred for the list of the HDS 
	 sources.} 
\label{source_feh}
\end{figure}

Spectroscopic [Fe/H] abundances based on HDS methodology were 
collected in a multistep process, whereby the additional data 
sets were iteratively examined for quality control. This 
process has led to re-instating sources that were priorly 
excluded and vice verse. Unfortunately, we had to exclude two 
sources \citep[][]{gilligan2021,sprague2022} due to excessive 
scatter around the [Fe/H] ridge established by the other sources. 
Examples on a well-fit subset and one of the excluded sets are  
shown in Fig.~\ref{source_feh}.  
%
%
Finally, the set of 17 sources was established, with $1279$ 
individual observations, often multiple visits to the 
same objects. Among the targets there were many RRc stars, 
several RRd, BL~Her and binary stars. These, together with 
faint ($V\gtrsim 15$~mag) objects (of any type) were excluded 
(altogether $104$ objects) from further analysis. The final 
inventory of $269$ RRab stars is shown in Table~\ref{hds_inv}.    

Concerning the data handling, the following points worth mentioning.
For the ``big survey'' data (LAMOST and GALAH) an upper error limit 
of $0.4$~dex was employed. After a considerable amount of testing, 
we decided to use simple averaging instead of robust averaging for 
the computation of the final metallicities for objects with multiple 
sources. This choice was justified because of the overall moderate 
number of multiple measurements for the individual objects. In merging 
the HDS and LDS data we considered the individual measurements in 
both sets to be equal, therefore, the final [Fe/H] was computed as a  
weighted average, i.e., [Fe/H]$=w$[Fe/H]$_{HDS}+(1-w)$[Fe/H]$_{LDS}$, 
where $w=N_{HDS}/(N_{HDS}+N_{LDS})$, and the $N$-s stand for the number 
of spectra. Except for the following two stars, all objects were 
treated equally. We did not use the HDS values of V0455~Oph, because 
there were differences up to $0.9$~dex among the four published values. 
For WW~Vir, the value of \cite{xiang2019} was an extreme outlier with 
respect of the values of \cite{luo2019a,luo2019b} and \cite{layden1994}, 
so, Xiang's et al. value was omitted.

%
%
\section{Light curves}
\label{app_B}
We opted to employ the long-term V-band time series photometry supplied 
by the ASAS and ASAS-SN surveys. These projects yield unique data sets 
fitting perfectly to our goals to derive: (i) reliable low-order phases 
for testing the Fourier-based [Fe/H] estimation; (ii) homogeneous sets 
of V magnitude averages for testing the $M_V$ dependence on [Fe/H]; 
(iii) proper decomposition of the light curves into the monomode and 
Blazhko components for goal (i) and for the assessment of the role of 
Blazhko phenomenon in the applications.        

For variables without an apparent Blazhko modulation, we employed 
standard least squares Fourier fit up to order $15$ (depending on 
the quality of the light curve). Outliers were iteratively clipped. 
For the Blazhko stars, first we performed a scan of the Blazhko 
period by using the following simple Fourier model, considering 
only the first order modulation side lobes
%
%
\begin{eqnarray}
\label{BL_fit}
V(t) &=& \sum_{i=1}^{m} A_i^{-}\sin(\omega_i^{-}t+\varphi_i^{-}) + 
A^{+}_i\sin(\omega_i^{+}t+\varphi_i^{+})\nonumber \\ 
     &+& \sum_{j=1}^{k} A_j\sin(\omega_jt+\varphi_j) \hspace{2mm} ,
\end{eqnarray}
where $m$ is the highest harmonics with side lobe frequencies 
$\omega_i^{-}$ and $\omega_i^{+}$. The Blazhko frequency is  
$\omega_B=\omega_i-\omega_i^{-}=\omega_i^{+}-\omega_i$, with 
$\omega_i$ being the $i^{th}$ harmonics of the fundamental 
frequency. We used $m=7$ in nearly all cases and constrained 
the monomode component to order $k\geq m$. Although the frequency 
spectra of Blazhko stars are, in general, more complicated 
than the simple model above, in a large majority of cases it 
yielded a good description of the data.  
 
For the Fourier phases and average magnitudes, in addition to 
ASAS and ASAS-SN, we considered also the data collected by 
\cite{jurcsik1996} based on classical individual stellar studies. 
The final set of $\varphi_{31}$ and the magnitude averages (V), 
result from the zero point shifted equal-weighted averages from 
all possible sources (by equal weighting all three sources). 
The zero point shifts -- \cite{jurcsik1996} minus source -- 
for $\varphi_{31}$ are $-0.015$ and $+0.015$ for ASAS and 
ASAS-SN, respectively. The shifts for (V) -- in the same order 
and same sense -- are as follows: $+0.016$ and $+0.039$ mag, 
with high respective standard deviations of $0.013$ and $0.037$. 

There are two stars that have not been used in any light 
curve-related tests. The bright Blazhko stars RR~Lyr and XZ~Cyg 
are only in the ASAS-SN database, but they are saturated, with  
XZ~Cyg having too few data points. TZ~Aur is monoperiodic, 
but older data were fractional, resulting questionable Fourier 
phases. Therefore, the star was not used from \cite{jurcsik1996}.   

%
%
\section{Distribution of the gravity coefficient}
\label{app_C}
\begin{figure}[h]
\centering
\includegraphics[width=0.40\textwidth]{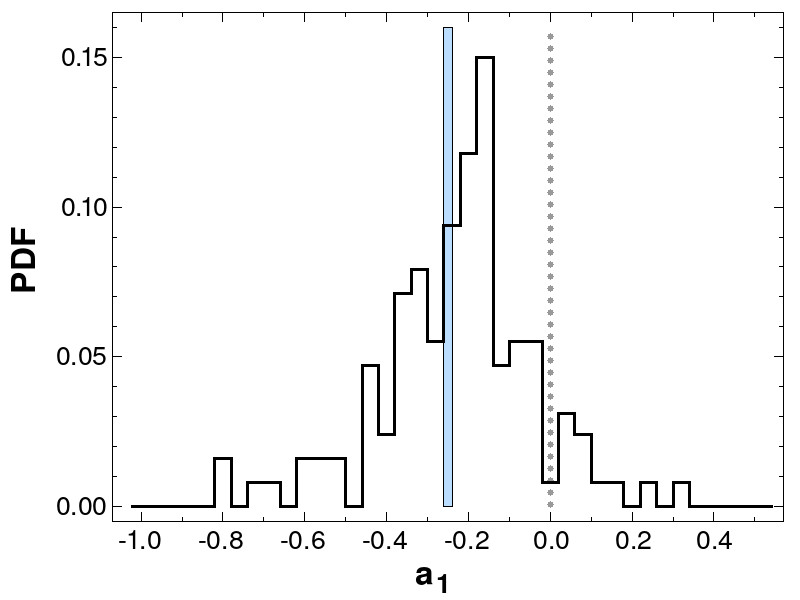}
\caption{Probability distribution function of the gravity 
         coefficient $a_1$ in the linear regression of 
	 $\Delta \log g$ to the published abundances (see 
	 Eq.~\ref{feh_logg_reg}). For reference, dotted 
	 and thick vertical lines, respectively, show the 
	 $a1=0.0$ and our final adopted value $a1=-0.25$.}  
\label{cgg_pdf}
\end{figure}

As noted in Sect.~\ref{sect:Cg}, individual cases might yield 
quite different values for the gravity factor $C_g$. To get an overall 
view on the statistics of the $C_g$ values, for each star, we fit the 
published [Fe/H] values by an equal-weighted least squares method 
using the following formula 
%
%
\begin{eqnarray}
\label{feh_logg_reg}
{\rm [Fe/H]}_{sp} = a_0 + a_1\Delta \log g \hspace{2mm} , 
\end{eqnarray}
where $\Delta \log g=\log g - \log g_{sp}$, with $\log g$ 
denoting the estimated static (Eq.~\ref{logp_logg}) and 
$\log g_{sp}$ the published spectroscopic gravities. 
In this equation coefficient $a_1$ corresponds to the object-dependent 
$C_g$ (with a negative sign, due to the re-arrengement of 
Eq.~\ref{feh_logg}). The distribution of \{$a_1$\} is shown in 
Fig.~\ref{cgg_pdf}. We see that the distribution is asymmetric 
around the zero slope and peaks close to our adopted value of 
$a_1=-0.25$. The plot shown was derived from the dataset obtained 
by selecting objects with more than two spectroscopic measurements. 
The dependence of the averages of \{$a_1$\} on the allowed minimum 
number of spectra is given in Table~\ref{tab_a1}. We used $3\sigma$ 
iterative clipping in each case to derive the averages and the 
corresponding RMS values. We note that the finally adopted value 
of $C_g$ weights also on the more negative value preferred by the 
Fourier method (see Fig.~\ref{cgg_scan}).    

%
%
%

%
%
\begin{table}[h]
\centering
\begin{minipage}{80mm}
\caption{Average gravity coefficient as a function of minimum number of spectra}
\label{tab_a1}
\scalebox{1.0}{
\begin{tabular}{cccrr}
\hline
 $Nsp_{min}$ & $\langle a_1 \rangle$ & $RMS(a_1)$ & $N$ & $N_{clip}$\\
\hline\hline
2 & -0.217 & 0.194 & 194 & 20\\
3 & -0.222 & 0.173 & 127 &  7\\
4 & -0.210 & 0.148 & 108 &  8\\
5 & -0.208 & 0.147 &  72 &  3\\ 
\hline
\end{tabular}}
\end{minipage}
\begin{flushleft}
\vspace{-5pt}
{\bf Notes:}
$Nsp_{min}$: minimum number of spectra per object; 
$\langle a_1 \rangle$: average of \{$a_1$\}; 
$RMS(a_1)$: RMS of \{$a_1$\}; 
$N$: Number of stars; 
$N_{clip}$: Number of stars clipped.      
\end{flushleft}
\end{table}
%

%
%
\section{Associated results}
\label{app_D}
Here we briefly summarize some of the consequences of the 
metallicity scale introduced in this paper. We focus on the 
residual distribution of the metallicity estimate based on 
the Fourier parameters, the classical metallicity--absolute 
V magnitude and the near infrared PLZ relations. 

\begin{figure}[h]
\centering
\includegraphics[width=0.40\textwidth]{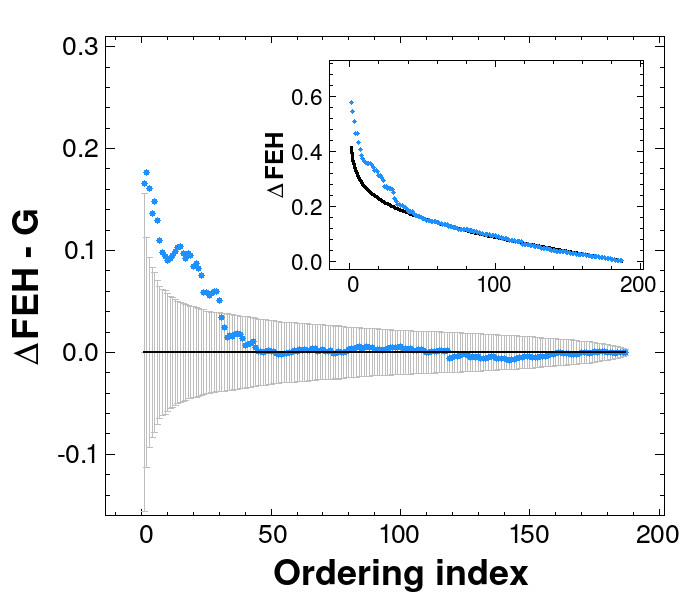}
\caption{Testing the $(P,\varphi_{31})\rightarrow{\rm FEH}$ fit 
        residuals.
	{\em Inset:} Ordered distribution (light blue) of the 
        residuals ($\Delta {\rm FEH}=|{\rm FEH_{obs}-FEH_{fit}}|$) 
	for set A of Table~\ref{data_31}. Black line shows the 
	distribution of the corresponding Gaussian, fitting the 
	core of the observed residuals. 
	{\em Main panel:} Difference between the observed and 
	the predicted Gaussian residuals. The $3\sigma$ ranges 
	for the Gaussian residuals are shown by the vertical 
	error bars.}  
\label{p_31_res}
\end{figure}

A general property of all metallicity sets studied in this 
paper is the non-Gaussian distribution of the subsets of stars 
lying in the gray zone of few sigma deviations with respect 
to the best linear fit of 
[Fe/H]$\sim a_0 + a_1P + a_2\varphi_{31}$. To demonstrate 
the size of the effect, we use set A of Table~\ref{data_31} 
(see also Fig.~\ref{p_31_fit} and Eq.~\ref{eq_p_31_fit} for the 
regression). The ordered residuals are shown in Fig.~\ref{p_31_res}. 
For some $80$\% of the sample, the distribution is within the 
$3\sigma$ range of the same size of sets following Gaussian 
distribution with a standard deviation of $\sigma_G=0.140$. 

As mentioned in Sect.~\ref{sect:jk96}, currently we have no 
explanation for this phenomenon. Once the outliers (both the 
non-fitting and non-Gaussian members -- 14 stars) are 
omitted, the resulting sample of 173 stars yields a fit 
RMS of $0.135$~dex and a core Gaussian distribution with 
$\sigma_G=0.127$~dex. It is important to recall, that, while 
some stars are suspect of light curve anomalies (e.g., IU~Car 
[period and light curve changes], SS~CVn [Blazhko with peculiar 
average light curve]) several stars, qualified as outliers 
(e.g., AL~CMi, X~Ari) have very accurate observed abundances, 
and no apparent light curve anomalies over several/many decades.  
%
%
%
%
%
%
%
\begin{figure}[h]
\centering
\includegraphics[width=0.40\textwidth]{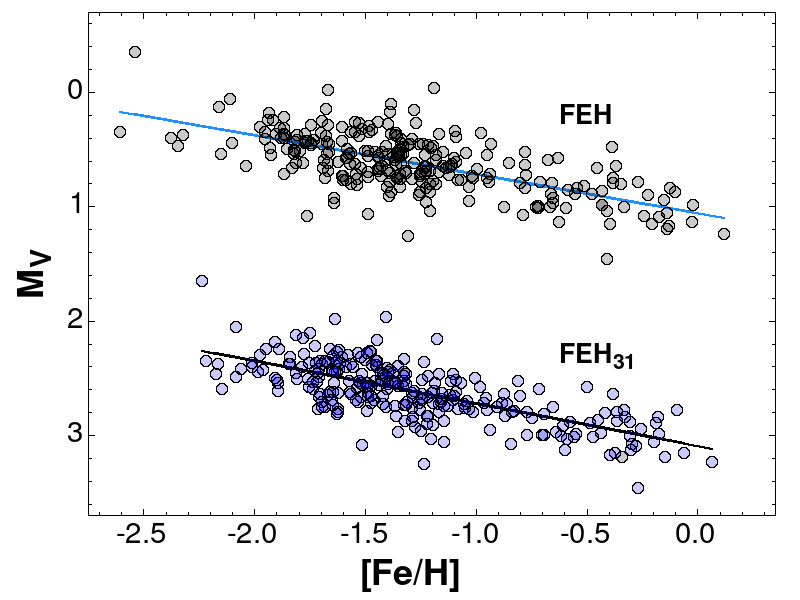}
\caption{Iron abundance vs V absolute magnitude, using a 
         subset of set B (see Table~\ref{data_31} and 
	 text for the selection of this subset). Lines are 
	 the resulting linear regressions. For better 
	 visibility, the set using Fourier-based abundances 
	 (FEH$_{31}$) has been shifted downward by $2$~mag 
	 with respect of the set using the mixture of  
	 gravity/source ZP-corrected HDS and LDS metallicities.}  
\label{feh_mv}
\end{figure}

The effect of the type of metallicity used in studying the 
metallicity-absolute magnitude relation is shown in Fig.~\ref{feh_mv}. 
We used the Gaia DR3 parallaxes \citep{lindegren2021} and reddenings 
by \cite{schlafly2011}. Set B of Table~\ref{data_31} was selected 
as a test base, but we employed cutoffs for $E(B-V)$ and for the 
relative parallax errors of $0.2$ and $0.06$, respectively. This 
had led to a sample of $264$ stars. To calculate the Fourier FEH, 
we used Eq.~\ref{eq_p_31_fit}. This yielded a residual standard 
deviation of $0.163$~mag (vs $0.184$ from the fit using direct FEH 
values). The difference is marginally significant (at the $2\sigma$ 
level). By using other samples, we lend on similar conclusions, 
often with a lower significance. As expected, the linear regressions 
are also very similar, yielding $M_V=1.059+0.341\times{\rm FEH}$ 
for the direct and $M_V=1.095+0.372\times{\rm FEH_{31}}$ for 
the Fourier-based FEH$_{31}$. We also examined the effect of 
gravity and source ZP corrections, and found similar low-significance 
preference for the corrected metallicities with respect to the 
uncorrected ones.  
%
%
%
\begin{figure}[h]
\centering
\includegraphics[width=0.45\textwidth]{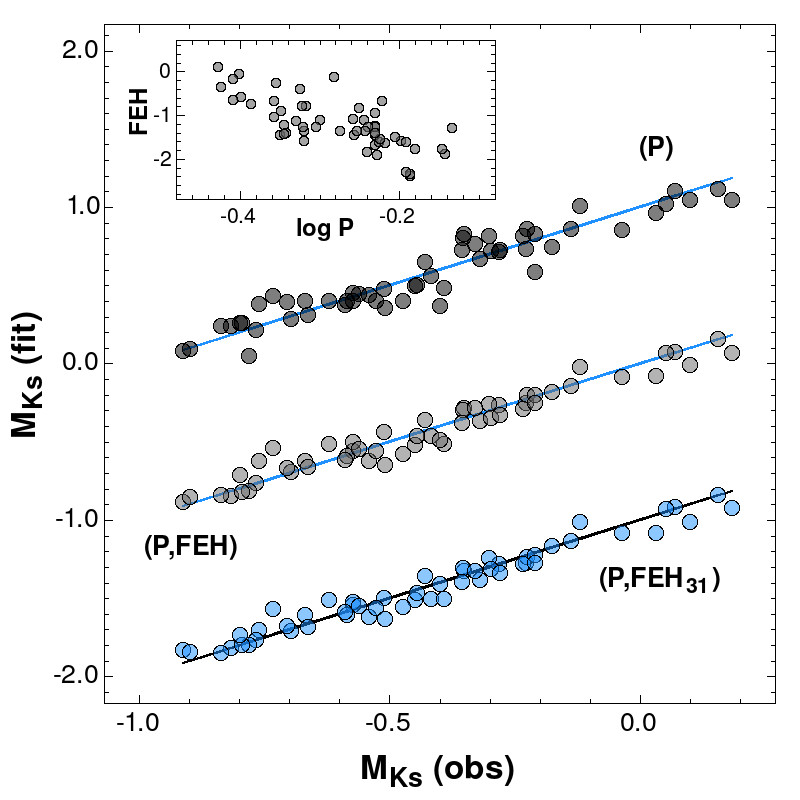}
\caption{Observed 2MASS absolute magnitude (Ks) vs predicted magnitudes 
         from various PLZ relations. The lines show the identity 
	 values, the labels indicate the formula type used in deriving 
	 the various fits. The inset shows the correlation between the 
	 two quantities fitted to the Ks magnitudes. See text for 
	 sample selection.}  
\label{feh_mk}
\end{figure}

Finally, in Fig.~\ref{feh_mk} we show the PL and PLZ relations in 
the near infrared. The goal of this test is to check: 
(a) the dependence on the metallicity used; 
(b) the significance of the metallicity dependence. 
We use the sample of $\sim 100$ stars with approximate average Ks 
magnitudes transformed from the unWISE fluxes \citep[][]{schlafly2019} 
as given by \cite{kovacs2021a}. 
After cross-matching with set B of Table~\ref{data_31} and 
applying $E(B-V)$ and relative parallax error cuts of 
$0.5$ and $0.03$, respectively, we ended up with a sample of $55$  
stars. As shown in Fig.~\ref{feh_mk}, we get increasingly better 
regressions from the single-parameter ($\log P$) fit to those 
utilizing the metallicities FEH and FEH$_{31}$. 
The latter (best) fit is significantly better than the single-parameter 
fit nearly at the level of $3\sigma$. On the other hand, we found 
basically no difference between using the original (published, not 
corrected) [Fe/H] and the gravity/source-corrected FEH. As shown 
in the inset of Fig.~\ref{feh_mk}, this is most probably attributed 
to the significant correlation between $\log P$ and [Fe/H] (the 
period dependence `takes over' if the available [Fe/H] becomes 
more noisy, yielding a fit of similar quality). Indeed, we obtained 
the following regressions:

\noindent 
$M_{Ks}=-1.4316-3.6331\times\log P$, 
 
\noindent 
$M_{Ks}=-0.9526 - 2.5676\times\log P + 0.1554\times{\rm FEH}$,
 
\noindent 
$M_{Ks}=-0.8874 - 2.4436\times\log P + 0.1899\times{\rm FEH_{31}}$, 

\noindent 
with $\sigma_{fit}=0.090$, $0.067$ and $0.060$, respectively.

\end{appendix}
\end{document}